\documentclass{PoS}
\usepackage{braket}
\usepackage{bm}
\usepackage{url}
\usepackage{amsmath}

\newcommand{\ta}{\mathtt{a}}
\newcommand{\tb}{\mathtt{b}}
\title{Tensor networks for gauge field theories}
\ShortTitle{TN for gauge field theories}

\author{\speaker{B. Buyens}$^{a}$, J. Haegeman$^{a}$, F. Verstraete$^{ab}$, K. Van Acoleyen$^a$\\
\llap{$^a$}Department of Physics and Astronomy, Ghent University, Krijgslaan 281, S9, 9000 Gent, Belgium \\
\llap{$^b$}Vienna Center for Quantum Science and Technology, Faculty of Physics,
University of Vienna, Boltzmanngasse 5, 1090 Vienna, Austria
\\ E-mail: \email{boye.buyens@ugent.be}, \email{jutho.haegeman@ugent.be}, \email{frank.verstaete@ugent.be}, \email{karel.vanacoleyen@ugent.be}}

\abstract{Over the last decade tensor network states (TNS) have emerged as a powerful tool for the study of quantum many body systems. The matrix product states (MPS) are one particular class of TNS and are used for the simulation of (1+1)-dimensional systems. In this proceeding we use MPS to determine the elementary excitations of the Schwinger model in the presence of an electric background field. We obtain an estimate for the value of the background field where the one-particle excitation with the largest energy becomes unstable and decays into two other elementary particles with smaller energy.}

\FullConference{The 33rd International Symposium on Lattice Field Theory\\
		14 -18 July 2015\\
		Kobe International Conference Center, Kobe, Japan*}

\begin{document}
\noindent\textbf{1. Introduction}\\
\noindent The exponential scaling of the Hilbert space with the number of particles or sites of a quantum many body system makes any attempt of solving a realistic system by using exact diagonalization impossible. Fortunately, the study of entanglement has shown that the low energy states of local gapped Hamiltonians live in a tiny corner of Hilbert space. This tiny corner corresponds to states for which the entanglement entropy of region scales with the boundary of that region instead of being extensive. Because they dominate the behavior of condensed matter systems at low temperatures, it makes sense to focus on the low energy states. Also for quantum field theories, independent of whether they are strongly coupled or weakly coupled, the low energy regime is of interest.
\nolinebreak
\\ \indent Tensor network states (TNS) \cite{Orus2013} are a variational class of states living in this tiny corner. Ideally, the number of parameters of these states is small and expectation values of local quantities can be computed efficiently in the number of sites in the system. In one spatial dimension the most famous example are the matrix product states (MPS). It is rigorously proven that they can efficiently approximate the low energy states of local gapped Hamiltonians \cite{Hastings2007}. The many successful simulations of many-body systems using MPS in the last decade showed that this result is not only of theoretical interest. Furthermore, as MPS are formulated in the Hamiltonian framework, they enable the difficult simulation of out-of-equilibrium physics \cite{Vidal1991,Banuls2009}. In the last years MPS have proven to be relevant for studying gauge theories, e.g. \cite{ZoharSB,Montangero2015}. In particular for the massive Schwinger model, QED$_2$ with one flavor, different groups considered MPS simulations, e.g. \cite{Byrnes2003,Buyens2013,Buyens2014,Buyens2015,Banuls2013, Banuls2015}. For higher dimensions different gauge invariant TNS constructions have also been developed \cite{Luca, Luca2, gaugingStates,Zohar2015} with some first numerical applications on simple gauge theories. 
\nolinebreak
\\ \indent Here we continue our research on the Schwinger model \cite{Buyens2013,Buyens2014,Buyens2015} by investigating the one-particle spectrum in the presence of an electric background electric field $g\alpha$. For $\alpha = 1/2$ something interesting happens. As the fermion mass increases there is a phase transition around $(m/g)_c \approx 0.33$ related to the spontaneous breaking of the CT-symmetry \cite{Byrnes2003,Coleman1976,Hamer}. The ground state is degenerate for $m/g > (m/g)_c$ and kinks `connecting' the two vacua arise. This is different from the spectrum in the case of a zero background electric field where the elementary particle spectrum consists of three or more stable particles for non-vanishing fermion mass \cite{Coleman1976}. In mass perturbation theory two stable particles survive for $\alpha < 1/4$ and only one for $1/4 < \alpha < 1/2$. As a new step in completing the phase diagram of the Schwinger model we will determine the elementary excitations for different values of $\alpha$ beyond mass perturbation theory. Earlier numerical studies on the elementary excitations in the non-perturbative regime of the Schwinger model exclusively focussed on the cases $\alpha = 0$ \cite{Byrnes2003, Banuls2013, Szyniszewski2013} and $\alpha = 1/2$ \cite{Byrnes2003}.  An overview of the low energy spectrum might also help in a better understanding of the dynamics induced by a quench in the form of an electric field. Indeed, in \cite{Buyens2014} we hinted that the behavior in the linear response regime can be understood by looking at the one-particle excitations of the Hamiltonian. Even beyond linear response, similar arguments explain the observations \cite{BuyensInPrep}.\\
\\
\noindent \textbf{2. Setup in the MPS-framework}\\
\noindent\textit{Hamiltonian.} The lattice Hamiltonian for the Schwinger model reads (see \cite{Buyens2015,Kogut1975} for details):
\begin{equation}\label{equationH} H= \frac{g}{2\sqrt{x}}\Biggl( \sum_{n \in \mathbb{Z}} \frac{{E}^2(n)}{g^2} + \sqrt{x}\frac{m}{g} \sum_{n \in \mathbb{Z}}(-1)^n\sigma_z(n) + x \sum_{n \in \mathbb{Z}}(\sigma^+ (n)e^{i\theta(n)}\sigma^-(n + 1) + h.c.)\biggl).\end{equation} 
Here we have introduced the parameter $x \equiv 1/(g^2a^2)$ with $a$ the lattice spacing, $m$ the fermion mass and $g$ the coupling constant. From now on we will work in units $g = 1$. The staggered fermions are traded for a spin system on the sites: $\sigma_z(n)\ket{s}_n= s \ket{s}_n (s=\pm 1), \sigma^{\pm}=(1/2)(\sigma_x\pm i \sigma_y)$. The gauge fields live on the links and we have the operators $\theta(n)=a g A_1(na)$ and their conjugate momenta $E(n)$ ($[\theta(n),E(n')]=i\delta_{n,n'}$), which correspond to the electric field. In an uniform electric background field $E = g\alpha$ in a compact formulation we have $E(n) = g(L(n) + \alpha)$ where $L(n)$ has  integer charge eigenvalues $p \in \mathbb{Z}$ and $e^{i\theta(n)}$ and $e^{-i\theta(n)}$ correspond to the ladders operators: $L(n)\ket{p}_n = p\ket{p}_n$, $e^{\pm i\theta(n)}\ket{p}_n = \ket{p\pm 1}_n, p \in \mathbb{Z}$. 

The key-feature of the Hamiltonian (\ref{equationH}) is that it has the gauge symmetry generated by $G(n) = L(n) - L(n-1) - (\sigma_z(n) + (-1)^n)/2$. A gauge theory differs from other theories by the fact that all physical states $\ket{\Phi}$ have to satisfy $G(n)\ket{\Phi} = 0, \forall n \in \mathbb{Z}$. Furthermore, the Hamiltonian is invariant under translations over an even number of sites. The fact that it is not invariant under a single translation originates from the staggered formulation. Finally, for $\alpha = 0$ and $\alpha = 1/2$ the Hamiltonian exhibits the CT symmetry which is the charge conjugation $C$ ($E \rightarrow - E$, $\sigma_z \rightarrow - \sigma_z$) added with a translation $T$ over one site. It is known that this $CT$ symmetry is only spontaneous broken for $\alpha = 1/2$ and $m/g \gtrsim 0.33$ \cite{Byrnes2003,Coleman1976}.\\
\\ \noindent\textit{Ground state and excitations.} In our approach site $n$ and link $n$ are blocked into an effective site with local Hilbert space spanned by the kets $\ket{\kappa}_n$ where ${\kappa} = (s,p) (s = \pm 1, p \in \mathbb{Z})$. In the thermodynamic limit ($N =  \infty$) we proposed in \cite{Buyens2015} the following ansatz for the ground state
\begin{subequations}\label{CTMPSgen}
\begin{equation}\label{CTMPS}\ket{\Psi(A)} = \sum_{\{ \kappa_n \}} v_L^\dagger \left(\prod_{n \in \mathbb{Z}}A_1^{\kappa_{2n-1}}A_2^{\kappa_{2n}}\right) v_R \ket{\bm{\kappa}}, \ket{\bm{\kappa}} = \{\ket{\kappa_{n}}\}_{n \in \mathbb{Z}},A_n^{\kappa} \in \mathbb{C}^{D_n \times D_{n+1}},v_L \in \mathbb{C}^{D_1}, v_R \in \mathbb{C}^{D_2}\end{equation}  
\begin{equation}\label{eqGIA}[A_n^{s,p}]_{(q,\alpha_q);(r,\beta_r)} =  \delta_{p, q + (s+(-1)^n)/2} \delta_{r,p}[\ta_n^{s,p}]_{\alpha_q, \beta_r}; q,r\in \mathbb{Z},\alpha_q = 1 \ldots D_q^n, \beta_r = 1\ldots D_r^{n+1}.\end{equation} 
\end{subequations}
One observes immediately that (\ref{CTMPS}) is manifest invariant under $T^2$ while the constraint (\ref{eqGIA}) imposes gauge invariance. The variational freedom of this state thus lies within the matrices $\ta_n^{s,p}$. For $\alpha = 0$ or $1/2$ one can impose $CT$ invariance by setting $A_2^{\kappa} = A_1^{{\kappa}^c}$ with $\kappa^c = (-s, -2\alpha - p)$ \cite{Buyens2013}. The optimal approximation within this class of states for the ground state is obtained using the time-dependent variational principle \cite{Buyens2015,Haegeman2011}.

For the elementary excitations with momentum $k$ we take the ansatz \cite{Haegeman2013a,Haegeman2013b}:
\begin{subequations}\label{excAnsatz}
\begin{equation} \ket{\Psi_k(B)} = \sum_{m \in \mathbb{Z}} e^{2ikn/\sqrt{x}} \sum_{\{\kappa_n\}} v_L^\dagger \left(\prod_{n \leq m}A_1^{\kappa_{2n-1}}A_2^{\kappa_{2n}}\right)B_k^{\kappa_{2m+1},\ldots,\kappa_{2(m+M)}} \left(\prod_{n > m+M}A_1^{\kappa_{2n-1}}A_2^{\kappa_{2n}}\right)v_R \ket{\bm{\kappa}},\end{equation} 
where $A_1$ and $A_2$ correspond to the ground state (\ref{CTMPSgen}) and gauge invariance is imposed by
\begin{equation}[B_k^{(s_1,p_1),\ldots,(s_{2M},p_{2M})}]_{(q,\alpha_q);(r,\beta_r)} = \left(\prod_{n = 2}^{2M}\delta_{p_n, p_{n-1}+(s + (-1)^n)/2}\right)\delta_{p_1,q + (s -1)/2}\delta_{p_{2M},r}[\tb_k^{p_1,s_1,\ldots,s_{2M}}]_{\alpha_q,\beta_r}.\end{equation}
\end{subequations}
For  $\alpha = 0$ and $\alpha = 1/2$ one can further constrain $B_k$ to classify the states according to their $CT$-number, see \cite{Buyens2013} for an example. In this case the excitations with $CT = -1$ are referred to as `vector' particles and excitations with $CT = 1$ are referred to as `scalar' particles. 

By minimizing $\braket{\Psi_k(\bar{B}_k) \vert H \vert \Psi_k(B_k)}/\braket{\Psi_k(\bar{B}_k) \vert \Psi_k(B_k)}$ with respect to $\tb_k$ one finds the optimal approximations $\ket{\Psi_k(B_k)}$ for the excitations. For sufficiently large bond dimension this ansatz should converge exponentially fast as $M$ increases to an elementary particle with momentum $k$ \cite{Haegeman2013a}. The speed of convergence depends on how far this excitation is separated from the other excitations in the same momentum sector (in units of the Lieb-Robinson velocity). Note that the computation time scales as $\mathcal{O}(4^M\max_{p} D_{p}^3)$ allowing only simulations for small $M$.\\

\begin{figure}[t]
\begin{tabular}{rr}
 \includegraphics[width=72mm]{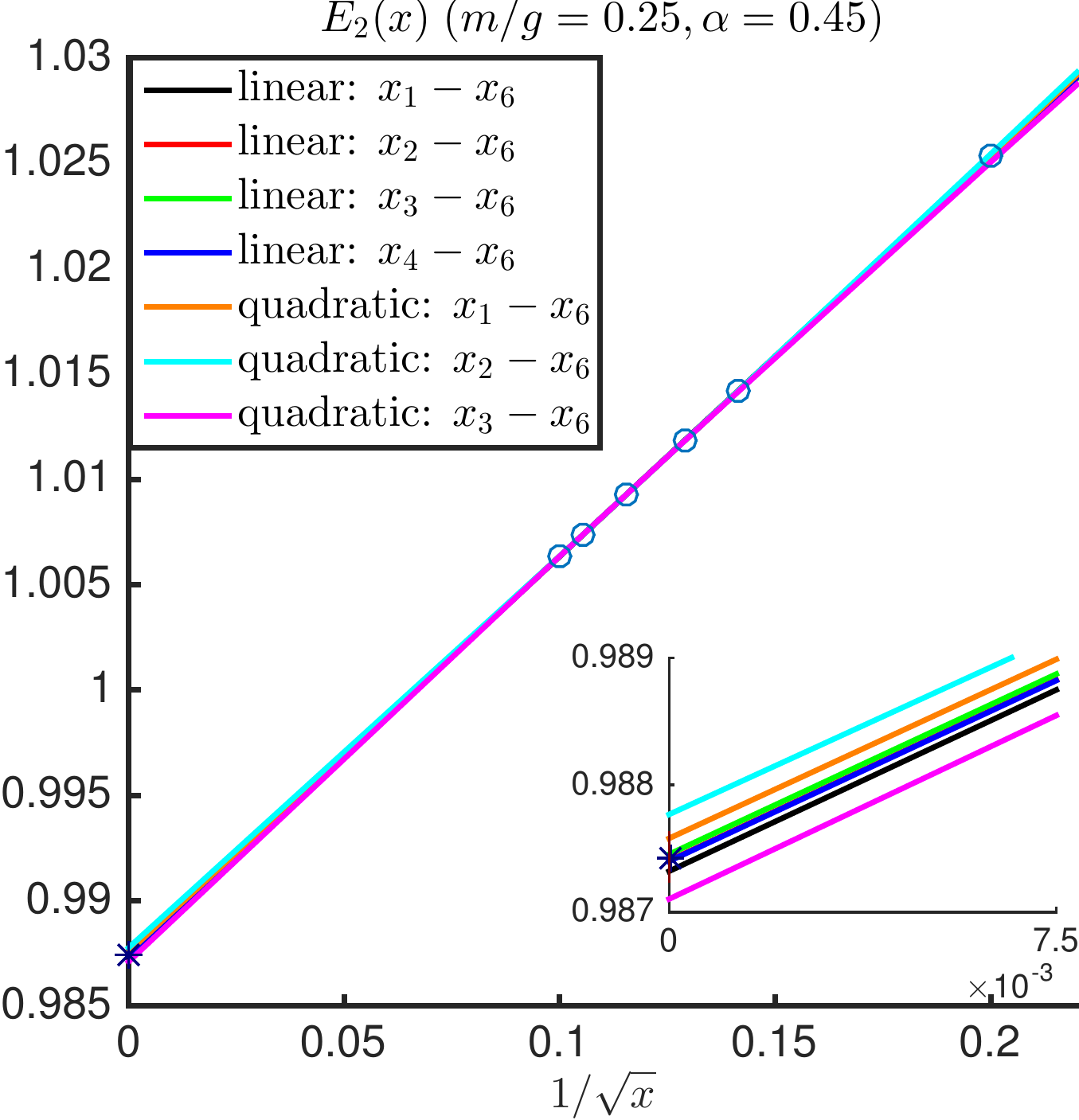}&\includegraphics[width=72mm]{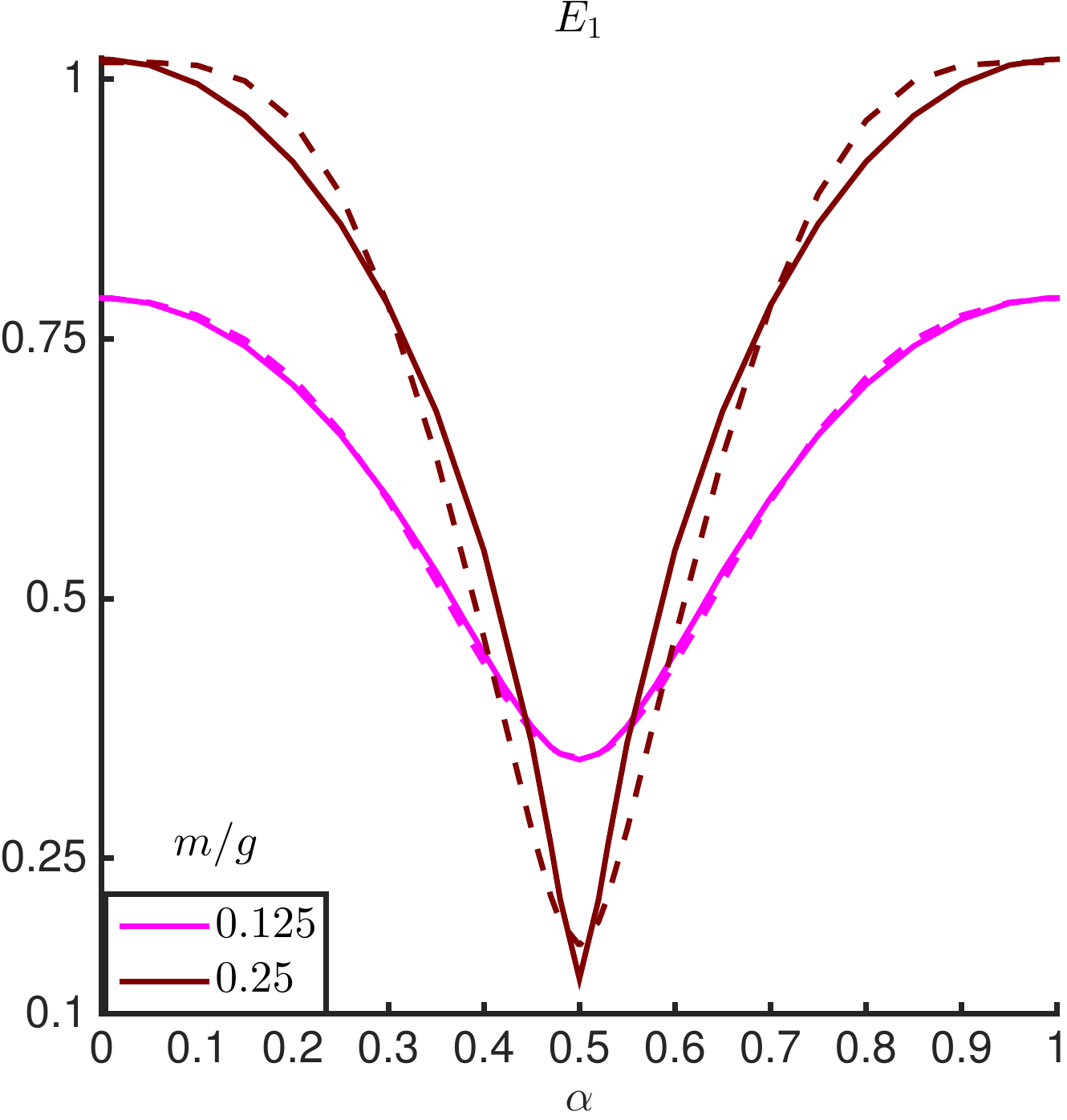}
 \end{tabular}
\caption{ Left (a): $m/g = 0.25, \alpha = 0.45$. Extrapolation of the energy $E_2$ of the second excited state to $x = \infty$. We perform several linear and quadratic fits in $1/\sqrt{x}$ through the points with $x = (x_1,\ldots, x_6) = (25,50,60,75,90,100)$ (see legend). Inset: our continuum estimate is the mean of all the fits. The error is the standard deviation. Right (b): Comparision of $E_1$ obtained by our numerical simulations (full line) with mass perturbation theory (dashed line).}
\label{fig:figure1}
\end{figure}

\noindent \textbf{3. Results and discussion}\\
\noindent   For a fixed value of $x$ we approximate the excited states using the ansatz (\ref{excAnsatz}) for $M = 2$. By comparing these energies with simulations for other values of the bond dimension $D_p$ and for $M = 1$ we obtain an error for truncating the bond dimension and truncating $M$.  Continuum estimates for the excitation energies are obtained similar as in \cite{Buyens2015,Banuls2013,Hamer}: we compute excitation energies for $x = 25, 50, 60, 75, 90, 100$ and perform linear fits in $1/\sqrt{x}$ through the points corresponding to the largest three, four, five and six x-values, see fig. \ref{fig:figure1} (a) . Furthermore we fit the points corresponding to the largest four, five and six $x-$values against a quadratic function in $1/\sqrt{x}$. All these fits give us an estimate of the energy in the continuum limit. To have some robustness against the choice of fitting interval and the fitting function we take the mean of all these energies as our final estimate. The standard deviation of this mean serves as an error on this value. In our simulations this standard deviation is not larger than of order $10^{-3}$ and dominates over the errors of truncating $D_p$ and truncating $M$. More accurate results can be achieved by taking larger $x-$values, although this will require a larger bond dimension and thus longer computation time. 

\begin{figure}[t]
\begin{tabular}{rr}
\includegraphics[width=72mm]{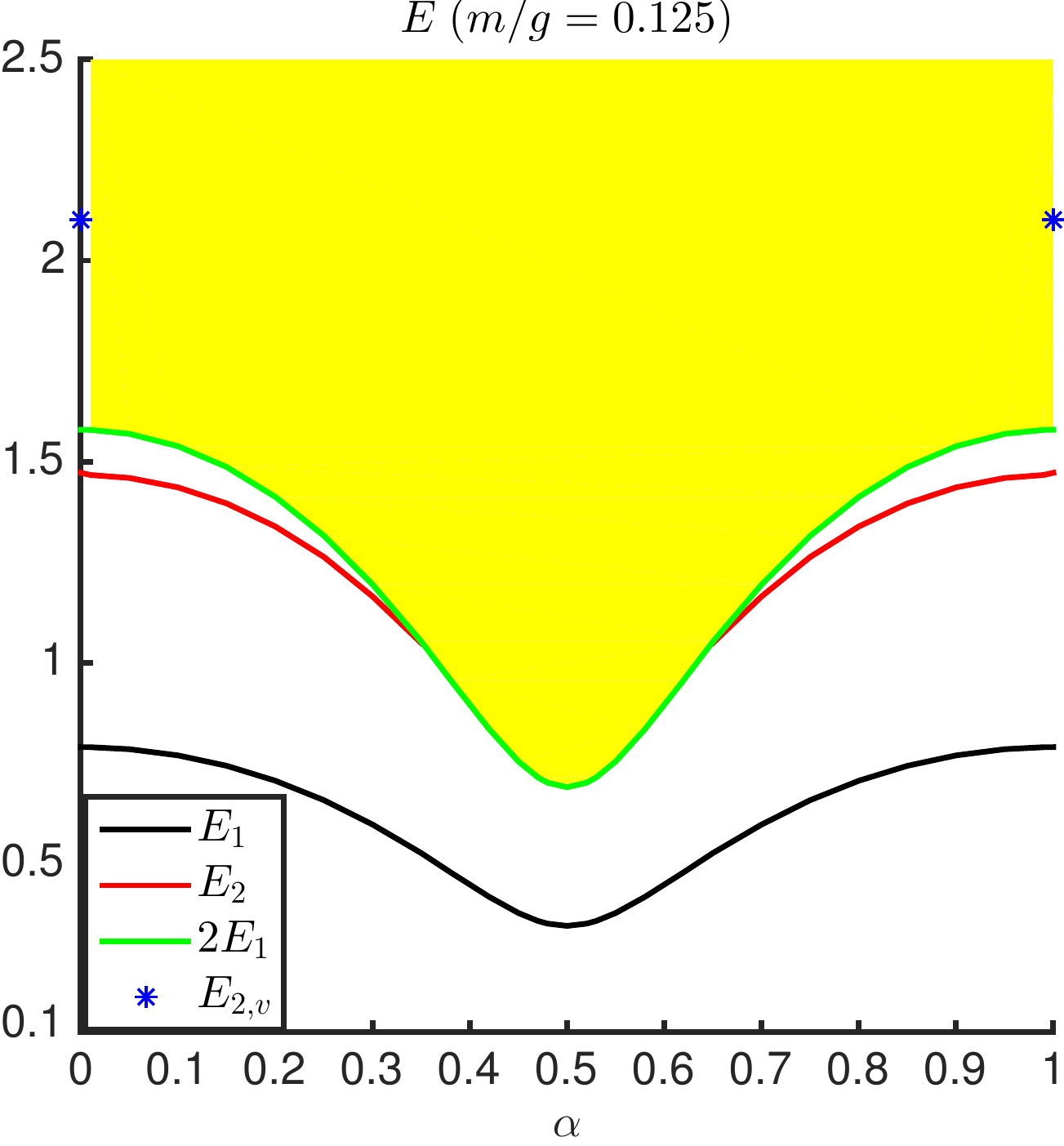}&\includegraphics[width=72mm]{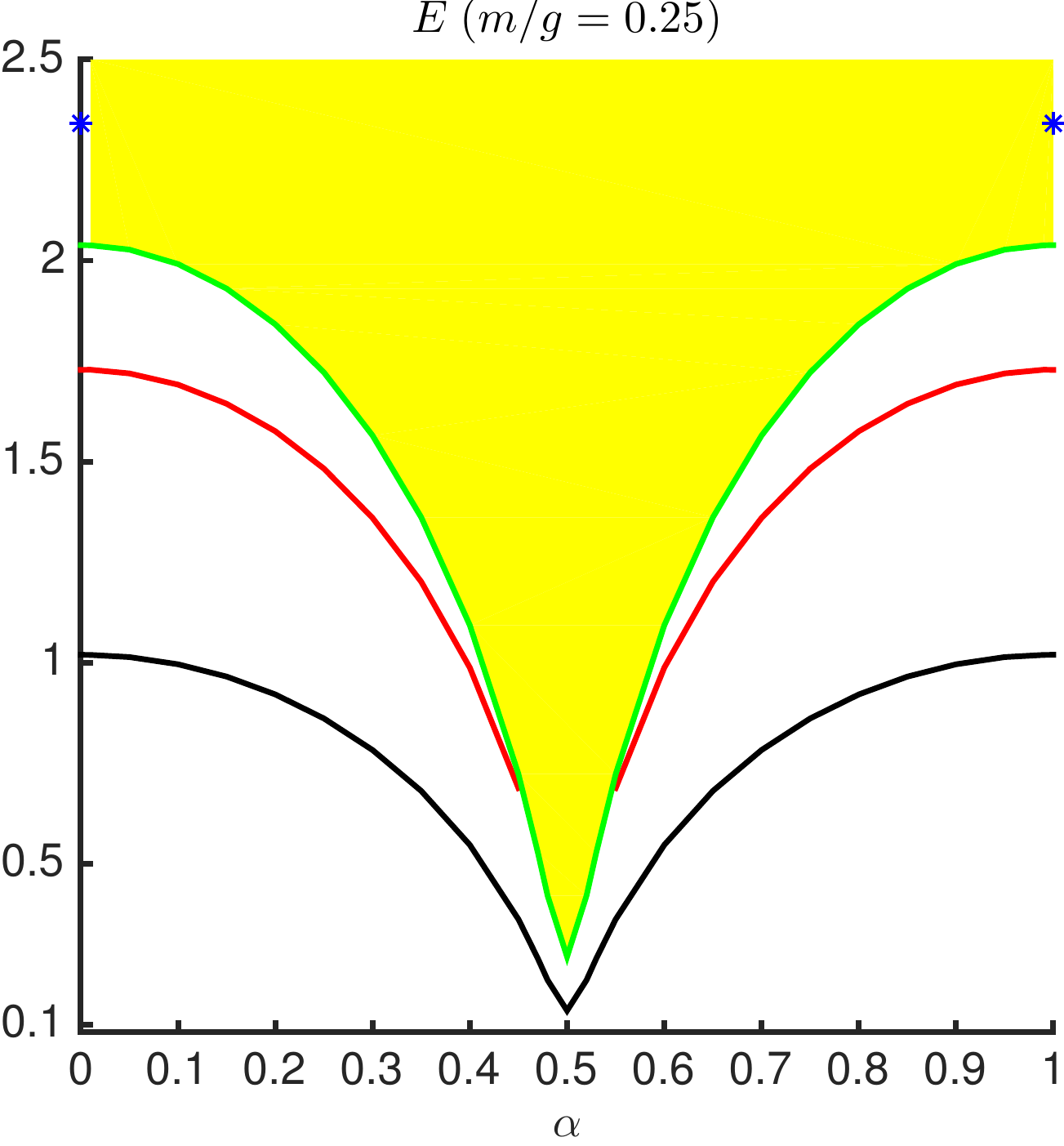}
\end{tabular}
\caption{Left (a): $m/g = 0.125.$ Energy of the elementary particles for different values of $\alpha$. For $\alpha \lesssim 0.35$ we detect two stable particles with energies $E_1$ and $E_2$. For $\alpha \gtrsim 0.38$ the particle with energy $E_2$ has disappeared in the continuum spectrum (yellow). Right (b): The same as (a) but for $m/g = 0.25$. Now the second particle is unstable for $\alpha \gtrsim 0.47$.}
\label{fig:figure2} 
\end{figure}

Note that physics is periodic in $\alpha$ with period $1$ and the excitations for $\alpha \in [1/2,1]$ can be obtained from the excitations for $\alpha \in [0,1/2]$ by a CT transformation. Therefore we can restrict our computation to $\alpha \in [0,1/2]$. Also, as the Schwinger model is a relativistic theory, the energy $E_k$ of a particle with momentum $k$ can be obtained from the energy $E_0$ of this excitation with momentum zero by the Einstein dispersion relation $E_k = \sqrt{k^2 + E_0}$. As a consequence we only need to compute the excitations with momentum zero. In fig \ref{fig:figure1} (b) we compare our numerical results for the energy $E_1$ of the first excited state with mass perturbation theory \cite{Adam1997}:
\begin{equation}\label{Adameq} E_1 = \mu_0 \sqrt{1 + 3.5621\frac{m}{\mu_0}\cos(2\pi\alpha) + 5.4807\left(\frac{m}{\mu_0}\right)^2 - 2.0933\left(\frac{m}{\mu_0}\right)^2\cos(4\pi\alpha)} + \mathcal{O}\left( \left[\frac{m}{\mu_0}\right]^3 \right).\end{equation}
where $\mu_0 = \frac{g}{\sqrt{\pi}}$. The plot shows that our numerical results converge towards (\ref{Adameq}) when $m/g \rightarrow 0$. \\
\\For $\alpha = 0$, we found in \cite{Buyens2013} two elementary excitations with $CT = -1$ and energy $E_{1,v}$ and $E_{2,v}$ and one elementary excitation with $CT = 1$ and energy $E_{1,s}$. For these energies we had $E_{2,v} < E_{1,v} + E_{1,s}$ but $E_{2,v} > 2E_{1,v}$, which means that the decay of $E_{2,v}$ into two elementary particles is only prevented by the $CT$ symmetry. When $0 < \alpha < 1/2$ the $CT$ symmetry is broken and this decay is no longer forbidden. This is indeed what we observe in the one-particle spectrum: for $\alpha > 0$ only the excitations with energy $E_1$ resp. $E_2$ corresponding to $E_{1,v}$ and $E_{1,s}$ for $\alpha \rightarrow 0$ remain stable, see fig. \ref{fig:figure2}. Furthermore, we observe that the binding energy $E_{bind}= 2E_1 - E_2$ decreases as $\alpha$ tends towards $1/2$. When the binding energy becomes small, the convergence rate of the ansatz (\ref{excAnsatz}) as a function of $M$ to the excited state with energy $E_2$ is rather slow. 

For $m/g = 0.125$, see fig \ref{fig:figure2} (a), the second particle is stable until $\alpha \lesssim 0.35$. For $\alpha = 0.38$ our estimates are $E_1 = 0.4784 (5)$ and $E_2 = 0.965 (2)$, indicating that the second excited state is unstable, $E_2 > 2E_1$. When $\alpha > 0.38$ we have $E_2(x) > 2E_1(x)$ for all the $x-$values we used. We conclude that there are two stable particles for $\alpha \lesssim 0.35$ and only one stable particle for $\alpha \gtrsim 0.38$.

For $m/g = 0.25$, see fig \ref{fig:figure2} (b), our estimates for the energy $E_2$ were unstable against variation of the bond dimension $D$ and $M$ for  $\alpha \geq 0.47$. The errors on $E_2$ are too large and prevent an extrapolation towards $x = \infty$. Nevertheless, in our simulations we have $E_2(x) < 2E_1(x)$ for $x = (25,50,60,75,90,100)$ and the fact that $E_2(x)$ decreases as the bond dimension and $M$ increase might suggest that this particle is still stable but with very small binding energy.  

For $\alpha = 1/2$ the ground state is $CT$ invariant. We computed the excitation energies with and without classifying the states according to their $CT-$number. In both cases, we found only one elementary particle. In the vector sector ($CT = -1$) all other states had energies that were larger than $3E_1$ and in the scalar sector ($CT = 1$) the energies were larger than $2E_1$. This corresponds to a theory with one stable particle. Therefore we estimate the value of the electric background field where the second elementary particle disappears to be larger than $0.47$ but smaller than $0.5$ for $m/g = 0.25$. 

In mass perturbation theory it is shown that there are two stable particles for $\alpha \leq 1/4$ and one stable particle for $1/4 < \alpha \leq 1/2$ \cite{Coleman1976}. Together with our results in the non-perturbative regime, we conclude that the value of $\alpha_c$, where one particle disappears, tends to $1/2$ for increasing mass of the fermions. Note however that the mass gap also decreases for larger $m/g$ and eventually the system becomes gapless for $m/g = (m/g)_c \approx 0.33$ and $\alpha = 1/2$ \cite{Byrnes2003}.  

In \cite{Byrnes2003} DMRG was used on a finite lattice to investigate the Schwinger model for $\alpha = 1/2$ and a degeneracy for the first elementary excitation was reported. Contrary, our simulations do not show any traces of that. 
\\
\\\textbf{4. Conclusion}\\
\noindent In this proceeding we continued the exploration of the Schwinger model as a testbed for MPS simulation of Hamiltonian lattice gauge theories. We investigated the elementary particle spectrum for a non-zero background electric field. For $m/g = 0.125, 0.25$ and small values of this background field we detected two stable particles, but as $\alpha \rightarrow 1/2$ one particle disappears in the continuum of the spectrum. This mechanism is best understood as the binding energy of the second excited state becoming too small to  be stable against a decay into two elementary particles with smaller energy. It would definitely be interesting to observe what happens closer to the critical point $(m/g, \alpha) = (0.33,1/2)$.

Looking further afield a logic step is the simulations of non-abelian lattice gauge theories in 1+1 and the simulations of higher dimensional gauge theories. The latter will be more challenging as the present algorithms for PEPS, the two-dimensional analogue of MPS, are restricted to relatively small bond dimension. Nevertheless, in the last years there has been some progress in PEPS algorithms \cite{Corboz2009, Jordan2008, Vanderstraeten2015b}. Furthermore, as this approach is free of any sign problem and enables real-time simulations, it is certainly worthwhile to further explore in this direction. \\
\\ \noindent \emph{Acknowledgements. }We acknowledge very interesting discussions with Laurens Vanderstraeten. This work is supported by an Odysseus grant from the FWO, a PhD-grant from the FWO (B.B), a post-doc grant from the FWO (J.H.), the FWF grants FoQuS and Vicom, the ERC grant QUERG and the EU grant SIQS.

\bibliographystyle{h-physrev}
\bibliography{skeleton}

\begin{thebibliography}{10}

\bibitem{Orus2013}
R.~{Or{\'u}s},
\newblock Annals of Physics {\bf 349}, 117 (2014).

\bibitem{Hastings2007}
M.~B. {Hastings},
\newblock Journal of Statistical Mechanics: Theory and Experiment {\bf 8}, 24
  (2007).

\bibitem{Vidal1991}
G.~Vidal,
\newblock Phys. Rev. Lett. {\bf 91}, 147902 (2003).

\bibitem{Banuls2009}
M.~C. Ba\~nuls, M.~B. Hastings, F.~Verstraete, and J.~I. Cirac,
\newblock Phys. Rev. Lett. {\bf 102}, 240603 (2009).

\bibitem{ZoharSB}
S.~{K{\"u}hn}, E.~{Zohar}, J.~I. {Cirac}, and M.~C. {Ba{\~n}uls},
\newblock Journal of High Energy Physics {\bf 7}, 130 (2015), arXiv 1505.04441.

\bibitem{Montangero2015}
T.~{Pichler}, M.~{Dalmonte}, E.~{Rico}, P.~{Zoller}, and S.~{Montangero},
\newblock (2015), arXiv 1505.04440.

\bibitem{Byrnes2003}
T.~Byrnes,
\newblock {\em \textit{Density Matrix Renormalization Group: A New Approach to
  Lattice Gauge Theory}} (University of New South Wales, 2003).

\bibitem{Buyens2013}
B.~{Buyens}, J.~{Haegeman}, K.~{Van Acoleyen}, H.~{Verschelde}, and
  F.~{Verstraete},
\newblock Physical Review Letters {\bf 113} (2014).

\bibitem{Buyens2014}
B.~{Buyens}, K.~{Van Acoleyen}, J.~{Haegeman}, and F.~{Verstraete},
\newblock (2014), arXiv 1411.0020.

\bibitem{Buyens2015}
B.~{Buyens}, J.~{Haegeman}, H.~{Verschelde}, F.~{Verstraete}, and K.~{Van
  Acoleyen},
\newblock (2015), arXiv 1509.00246.

\bibitem{Banuls2013}
M.~C. {Ba{\~n}uls}, K.~{Cichy}, J.~I. {Cirac}, and K.~{Jansen},
\newblock Journal of High Energy Physics {\bf 11}, 158 (2013).

\bibitem{Banuls2015}
M.~C. {Ba{\~n}uls}, K.~{Cichy}, J.~I. {Cirac}, K.~{Jansen}, and H.~{Saito},
\newblock Phys. Rev. D {\bf 92}, 034519 (2015).

\bibitem{Luca}
L.~{Tagliacozzo} and G.~{Vidal},
\newblock Phys. Rev. B {\bf 83}, 115127 (2011).

\bibitem{Luca2}
L.~{Tagliacozzo}, A.~{Celi}, and M.~{Lewenstein},
\newblock Physical Review X {\bf 4}, 041024 (2014).

\bibitem{gaugingStates}
J.~{Haegeman}, K.~{Van Acoleyen}, N.~{Schuch}, J.~I. {Cirac}, and
  F.~{Verstraete},
\newblock Physical Review X {\bf 5}, 011024 (2015).

\bibitem{Zohar2015}
E.~{Zohar}, M.~{Burrello}, T.~{Wahl}, and J.~I. {Cirac},
\newblock (2015), arXiv 1507.08837.

\bibitem{Coleman1976}
S.~R. Coleman,
\newblock Annals Phys. {\bf 101} (1976).

\bibitem{Hamer}
C.~Hamer, J.~Kogut, D.~Crewther, and M.~Mazzolini,
\newblock Nuclear Physics B {\bf 208}, 413  (1982).

\bibitem{Szyniszewski2013}
M.~{Szyniszewski},
\newblock (2013), arXiv 1303.0441.

\bibitem{BuyensInPrep}
B.~{Buyens}, J.~{Haegeman}, H.~{Verschelde}, F.~{Verstraete}, and K.~{Van
  Acoleyen},
\newblock in preparation .

\bibitem{Kogut1975}
J.~Kogut and L.~Susskind,
\newblock Phys. Rev. D {\bf 11} (1975).

\bibitem{Haegeman2011}
J.~{Haegeman} {\em et~al.},
\newblock Physical Review Letters {\bf 107} (2011).

\bibitem{Haegeman2013a}
J.~{Haegeman} {\em et~al.},
\newblock Physical Review Letters {\bf 111} (2013).

\bibitem{Haegeman2013b}
J.~{Haegeman}, T.~J. {Osborne}, and F.~{Verstraete},
\newblock Physical Review B {\bf 88} (2013).

\bibitem{Adam1997}
C.~{Adam},
\newblock Annals of Physics {\bf 259}, 1 (1997), hep-th/9704064.

\bibitem{Corboz2009}
P.~{Corboz}, R.~{Or{\'u}s}, B.~{Bauer}, and G.~{Vidal},
\newblock Physical Review B {\bf 81}, 165104 (2010), 0912.0646.

\bibitem{Jordan2008}
J.~{Jordan}, R.~{Or{\'u}s}, G.~{Vidal}, F.~{Verstraete}, and J.~I. {Cirac},
\newblock Physical Review Letters {\bf 101}, 250602 (2008), cond-mat/0703788.

\bibitem{Vanderstraeten2015b}
L.~{Vanderstraeten}, F.~{Verstraete}, and J.~{Haegeman},
\newblock (2015), arXiv 1507.02151.

\end{thebibliography}

\end{document}